\documentclass[12pt]{article}

\usepackage{amsmath,amsfonts,amssymb}

\def\bq{\mathbf{q}}
\def\bp{\mathbf{p}}

\def\hL{\hat{L}}

\def\be{\begin{equation}}

\def\ee{\end{equation}}

\def\bea{\begin{eqnarray}}

\def\eea{\end{eqnarray}}

\def\tr{\mathrm{tr}\, }

\def\tr{\mathrm{Tr}}

\newcommand{\hrho}{\hat{\rho}}

\newcommand{\hPi}{\hat{\Pi}}

\def\bra #1{\left<#1\right|}
\def\ket #1{\left|#1\right>}

\def\pb #1{\left\{#1\right\}}

\begin{document}

\begin{titlepage}

\vskip 0.4 cm

\begin{center}
{\Large{ \bf Note About Koopman-von Neumann Theory and Density Matrix}}

\vspace{1em}  Josef Kluso\v{n}$\,^1$
\footnote{Email address:
 klu@physics.muni.cz}\\
\vspace{1em} $^1$\textit{Department of Theoretical Physics and
Astrophysics, Faculty
of Science,\\
Masaryk University, Kotl\'a\v{r}sk\'a 2, 611 37, Brno, Czech Republic}\\

%
%

\vskip 0.8cm

\end{center}
 
\begin{abstract}
In this short note we study Koopman-von Neumann theory for N-particle system. We argue that it is natural to identify classical $N-$particle distribution function 
as diagonal form of density matrix operator in coordinate representation. We also determine generalized BBGKY hierarchy for reduced  density matrix in coordinate representation.

\end{abstract}

\bigskip

\end{titlepage}

\newpage

\section{Introduction}
Standard formulations of classical and quantum mechanics use completely different languages. In classical mechanics system of $N-$particles in three dimensional space is described by point in $6N-$dimensional phase space with exact values of coordinates and momenta. Observables are then defined as functions on this phase space. In case of quantum mechanics  system is described by wave function that depends on coordinates only and observables are represented by operators that act on it. These mathematical descriptions are completely different  and the transition from quantum mechanics description to the classical one is not completely clear. On possibility, how to relate classical with quantum description, is operational formulation of classical mechanics which was proposed by Koopman and von Neumann (KvN) \cite{Koopman,Neumann} where the classical mechanics is formulated on the  basis of Hilbert spaces. In this formulation we describe classical system with distribution function 
that is defined on the whole phase space and which determines probability that given system can be found in concrete point of phase space.
 KvN theory postulates that one particle distribution function can be defined in the same way as in quantum mechanics so that it is given as s product of classical wave function and its complex conjugate. Then  the time evolution of wave function is postulated
 to be generated by   Liouville operator so that 
wave function evolves in the same way as corresponding distribution function.  It is also important to stress that  classical wave function  depends on coordinate and momenta simultaneously since distribution function determines probability that we find system in given point of phase space while in quantum mechanics wave function depends on coordinates only.
 
KvN theory is remarkable attempt to find common language for classical and quantum physics, for some recent interesting papers, see \cite{Amin:2025stt,Amin:2026kai,Sen:2025urs,Sen:2022vig,Klein:2020djt,Gozzi:2003sh,Deotto:2002hy,Mauro:2001rm,Bondar:2013aqf}. 
$N-$particle system is described by $N-$particle distribution function whose time evolution is determined by  Liouville operator. 
 One characteristic property of the description of system with the help of distribution function is that the reduced distribution functions that describe dynamics of some subsystem are given by integration of $N-$particle distribution function over complement of the phase-space of subsystem we are interested in. Physically it meas that since we are not interested in the dynamics of the rest of particles we simply integrate corresponding probabilities
\cite{kintheory}. 

The question is how similar  procedure can be realized in KvN theory. 
In this article we argue that this procedure can be naturally included in 
KvN theory when we construct density matrix operator from 
the wave function that obeys Liouville equation. Then $N-$point distribution  function 
as we know from statistical physics 
corresponds to 
diagonal elements of this density matrix operator in coordinate representation. This construction allows to define classical density matrix in abstract operator form which is close to the idea of operational formulation of classical mechanics
\cite{Koopman,Neumann}.  We also determine equation of motion for density matrix
that has the form of the commutator between Liouville operator and density matrix. 
We further show that reduced 
 density matrix operators  are naturally determined as trace of $N-$particle density matrix over subspaces of Hilbert space which are complement
of the space we are interested in. In other words this procedure corresponds to the standard definition of reduced density matrix in quantum mechanics. Then finally we determine equations of motion for lower dimensional density matrix and we find that
they have the form of
BBGKY (Bogoliubov-Born-Green-Kirkwood-Yvon) hierarchy 
\cite{Yvon,Bogoliubov,Kirkwood,Born}.
This paper is organized as follows. In the next section (\ref{second}) we give a brief introduction to KvN theory on an example of single particle. Then in section (\ref{third}) 
we consider KvN theory for $N-$particles and we define   matrix density operator
in operational formulation of classical mechanics and study its properties.  Finally in conclusion (\ref{fourth}) we outline our results and suggest possible 
extension of this work.

\section{Introduction to Koopman-von Neumann theory}\label{second}
In this section we give  a brief  introduction to Koopman-von Neumann theory,
for extended review, see \cite{Mauro:2003sno}. Let us start
with one-particle distribution function $\rho_1(p,q,t)$ that gives probability that particle can be found in the  specific point of phase space $(p,q)$. For simplicity we consider one-dimensional 
motion so that corresponding phase space is two dimensional.
 It is well known that $\rho_1$ evolves in time according to the Liouville equation 
\begin{equation}\label{Liouville1}
	\frac{\partial \rho_1}{\partial t}+\pb{\rho,H_1}=
\frac{\partial \rho_1}{\partial t}+\left(\frac{\partial \rho_1}{\partial q}
\frac{\partial H_1}{\partial p}-\frac{\partial \rho_1}{\partial p}\frac{\partial H_1}{
\partial q}\right)=0 \ , 
\end{equation}
where $H_1$ is one particle Hamiltonian.

It is possible to rewrite the equation (\ref{Liouville1}) into operator-like form 
\begin{equation}
i\partial_t \rho_1=\hL_1\rho_1 \ , \quad 
\hL_1=i\left(\frac{\partial H_1}{\partial q}\frac{\partial }{\partial p}-
\frac{\partial H_1}{\partial p}\frac{\partial }{\partial q}\right) \ .
\end{equation}
Koopman-von Neumman theory is based on a simple presumption that   $\rho_1(p,q)$  is given as a product of complex function $\psi_1(p,q)$ and its complex conjugate 
\begin{equation}
	\rho_1(p,q)=\psi_1(p,q)\psi_1^*(p,q) \ , 
\end{equation}
where $\psi_1(q,p)$ is a complex and square integrable wave function. In KvN theory it is postulated that $\psi_1(q,p)$ obeys the same equation as $\rho_1$
\begin{equation}\label{parttpsi}
i\partial_t\psi_1=\hL_1\psi_1 \ . 
\end{equation}
Note that complex conjugate function $\psi_1^*$ obeys the same equation as $\psi_1$
\footnote{More general possibility was recently analyzed in 
\cite{Amin:2025stt}.}
\begin{equation}
	i\partial_t\psi_1^*=\hL_1\psi_1^* \  
\end{equation}	
using an important result $\hL_1^*=-\hL_1$. 
It is natural to interpret (\ref{parttpsi}) as an analogue of Schr\"{o}dinger
equation in classical mechanics that determines evolution of vectors
$\psi_1$  in the
Hilbert space of classical mechanics. These vectors obey normalization condition 
\begin{equation}
	\int dp dq \psi_1^*(p,q)\psi_1(p,q)=1 
\end{equation}	
that follows from the fact that $\rho_1$, as probability density, obeys the equation 
\begin{equation}
	\int dp dq\rho_1(p,q)=1 \ . 
\end{equation}
 It is crucial
that $\psi_1$ depends simultaneously on $p$ and $q$ since  particle has definite value of $q$ and $p$ in classical mechanics. 
 Consequently the scalar product of two functions
$\phi(q,p)$ and $\psi(q,p)$ on classical Hilbert space is defined as
\begin{equation}
\left<\phi|\psi\right>=\int dp dq \phi^*(p,q)\psi(p,q) \ ,
\end{equation}
where we use bra-ket notations for vectors and dual forms in Hilbert space where as in quantum mechanics state $\ket{\psi}$ is represented by wave function $\psi$ which can be naturally defined as
\begin{equation}
	\psi(p,q)=\left<q,q|\psi\right> \ ,
\end{equation}
where $\bra{q,p}$ is dual to $\ket{q,p}$ which is eigenvector of $\hat{p}$ and $\hat{q}$ operators 
\begin{equation}
	\hat{q}\ket{q,p}=q\ket{q,p} \ , \quad 
	\hat{p}\ket{q,p}=p\ket{q,p} \ . 
\end{equation}
Then $\ket{q,p}$ form continuous basis of classical Hilbert space 
so that as in quantum mechanics we can define partition of unity in the form 
\begin{equation}
	\hat{I}=\int dp dq\ket{q,p}\bra{q,p}
\end{equation}
and hence any state $\ket{\psi}$ can be written as
\begin{equation}
	\ket{\psi}=\int dp dq \psi(q,p)\ket{q,p} \ .
\end{equation}
Having defined inner product in Hilbert space we can calculate
\begin{eqnarray}
&&\bra{\phi}\hL_1\ket{\psi}=
\int dp dq \phi^* i\left(\frac{\partial H_1}{\partial q}\frac{\partial }{\partial p}-
\frac{\partial H_1}{\partial p}\frac{\partial }{\partial q}\right)\psi=\nonumber \\
&&-i\int dp dq (\frac{\partial}{\partial p}\left[\frac{\partial H_1}{\partial q}\phi^*\right]\psi-\frac{\partial}{\partial q}\left[\frac{\partial H_1}{\partial p}\phi^*\right]\psi)=\nonumber \\
&&=-\int dp dq i(\frac{\partial \phi^*}{\partial p}\frac{\partial H_1}{\partial q}-\frac{\partial \phi^*}{\partial q}\frac{\partial H_1}{\partial p}\psi)=\nonumber \\
&&=\int dp dq [\hL_1\phi]^*\psi=\left<\hL_1\phi|\psi\right> \ , \nonumber \\
\end{eqnarray}
using the fact that $\frac{\partial^2 H_1}{\partial p \partial q}=\frac{\partial^2 H_1}{\partial p\partial q}$. On the other hand since Hermitean conjugate operator is defined as
\begin{equation}
	\bra{\phi}\hL_1\ket{\psi}=\left<\hL^\dag \phi|\psi\right>
\end{equation}
we obtain comparing these two equations an important result
\begin{equation}\label{Ldag}
	\hL^\dag_1=\hL_1 \ . 
\end{equation}
In this section we gave a brief review of KvN theory for one-particle
distribution function. We saw that classical mechanics can be written in operational form which closely resembles formalism of quantum mechanics. In the next section we argue that this correspondence can be put even further when we consider $N-$particle
system.
 
\section{Koopman-von Neumann theory for $N-$particle system and Density Matrix Operator}\label{third}
 Fundamental ontology of $N-$particle classical mechanics is given by  positions and momenta of $N-$particles together with corresponding $N-$particle Hamiltonian $H_N$. State of the system is described by point in $6N-$dimensional phase space and it evolves according to Hamilton equations. Before we proceed further let us introduce following notation. Every particle is localized in three dimensional space with coordinates 
 $\bq_i=(q^1_i,q^2_i,q^3_i)$ and  corresponding momenta
$\bp_i=(p^1_i,p^2_i,p^3_i)$, where $i=1,\dots,N$ label individual particle. For simplicity of notation we use index $\Pi_i$ 
that label position of and momenta of  $i-$th particle 
so that  $\Pi_i=(\bq_i,\bp_i)$.

Let us consider again system of $N-$particles whose time evolution corresponds to trajectory in $6N-$dimensional phase space. 
On the other hand in it is possible  in the same way as in one particle case proceed to the field description of this system 
when we introduce $N-$particle distribution function $\rho_N(\Pi_1,\dots,\Pi_N)$ that gives a probability that the first particle is localized at the point $\Pi_1$ the second one at the point $\Pi_2$ and so on. In other words we perform step from the description of the whole system as single trajectory in the phase space to to the field description using $\rho_N(\Pi_1,\dots,\Pi_N)$ as a field defined on phase space.  Further $\rho_N$ obeys $N-$dimensional Liouville equation
\begin{equation}
	\frac{\partial \rho_N}{\partial t}+\pb{\rho_N,H_N}=0
\end{equation}
or explicitly
\begin{equation}
\frac{\partial \rho_N}{\partial t}+ \sum_{i=1}^N\left(\frac{\partial \rho_N}{\partial \bq^i}\cdot
\frac{\partial H_N}{\partial \bp_i}-\frac{\partial \rho_N}{\partial \bp_i}\cdot\frac{\partial H_N}
{\partial \bq_i}\right)=0\  . 
\end{equation}
As in previous section we write this equation into 
 operator like form 
\begin{equation}
	i\frac{\partial \rho_N}{\partial t}=\hL_N(\Pi)\rho_N \ , 
\end{equation}
where we introduced  $N-$particle Liouville operator $\hat{L}_N(\Pi)$ as 
\begin{equation}
	\hL_N(\Pi)=i
\sum_{i=1}^N\left(\frac{\partial H_N}
{\partial \bq_i}\cdot\frac{\partial }{\partial \bp_i}-
\frac{\partial H_N}{\partial \bp_i}\cdot\frac{\partial }{\partial \bq^i}\right)	 \ , 
\end{equation}
where we write explicit dependence of the Liouville operator $\hL_N(\Pi)$ on $\Pi$ in order to stress that it is defined
in coordinate representation.  Further,  $N-$particle Hamiltonian has the form 
\begin{equation}\label{HN}
	H_N=\sum_{i=1}^N \frac{1}{2m}\bp_i\cdot \bp_i+\frac{1}{2}\sum_{i=1}^N
	\sum_{j=1}^N V_{ij}(\bq_i,\bq_j) \ ,
\end{equation}
where for simplicity we ignore situation when the particles
interact with external field which will be represented by potential $V_{ex}(\bq_i)$.
Following single particle case discussed in previous section 
we presume  that $N-$particle
distribution function can be written as 
\begin{equation}\label{rhoNdef}
	\rho_N=\psi_N \psi_N^* \ , 
\end{equation}
where $\psi_N$ and $\psi_N^*$ obey equations
\begin{equation}\label{parttpsiN}
	i\partial_t\psi_N=\hL_N(\Pi)\psi_N \ , \quad i\partial_t\psi_N^*=\hL_N(\Pi)\psi_N^*  \ .
\end{equation}
Since our goal is to find operator formulation of classical
mechanics which is very close to the quantum mechanics 
 we would like to argue that (\ref{rhoNdef}) is coordinate
 representation of diagonal form of density matrix operator
 which we denote as $\hrho_N$. The reason for this proposal is
hidden in the relation between $N-$particle distribution functions and reduced ones. 
 Before we proceed to the explicit definition of $n-$point function let
us define $d\Pi_i$ as 
\begin{equation}
d\Pi_i=d^3\bq_i d^3\bp_i 
\end{equation}
so that volume element on $6N-$dimensional phase space is equal to 
\begin{equation}
	\prod_{i=1}^N d\Pi_i \ . 
\end{equation}
Note that $\rho_N$ is normalized as 
\begin{equation}
	\int 	\prod_{i=1}^N d\Pi_i \rho_N=1 \ . 
\end{equation}
Now we define  $n-$particle distribution function as \cite{kintheory}
\begin{equation}
	\rho_n(\Pi_1,\dots,\Pi_n)=
	\int \prod_{i=n+1}^N 
d\Pi_i \rho_N(\Pi_1,\dots,\Pi_N) \ . 
\end{equation}
Clearly this is natural definition of $N-$particle wave function in case of classical mechanics. However the question is
how this definition can be applied in operator description of classical mechanics. If we accept an existence of matrix density operator $\hrho_N$ we can very easily define reduced density matrix operator. Note that this procedure can be performed even in abstract notation without restriction to the coordinate representation. To do this 
let us introduce \emph{bra} and \emph{ket} vectors so that 
\begin{equation}
	\psi_N=\left<\Pi(N)|\psi_N\right> \ , \psi_N^*=\left<\psi_N|\Pi(N)\right> \ , 
\end{equation}
where we defined coordinate basis of Hilbert space 
as  
\begin{equation}
	\ket{\Pi(N)}=\ket{\Pi_1}\otimes \ket{\Pi_2}\otimes \dots \otimes \ket{\Pi_N} \ , 
\end{equation}
where $\otimes$ is tensor product and $\ket{\Pi_i}$ is eigenvector of operator $\hPi_i$ which is defined by ordinary multiplication 
\begin{equation}
	\hPi_i\ket{\Pi_j}=\delta_{ij}\Pi_i\ket{\Pi_i} \ . 
\end{equation}
Note that then we can rewrite the equations 
in (\ref{parttpsiN}) into manifestly basis independent form
\begin{equation}\label{Liouabst}
i\frac{d\ket{\psi_N}}{dt}=\hL_N\ket{\psi_N} \ , \quad 
i\frac{d\bra{\psi_N}}{dt}=-\bra{\psi_N}\hL_N \ .
\end{equation}
First equation in (\ref{parttpsiN}) can be easily derived from 
(\ref{Liouabst}) 
 when we multiply first equation in
(\ref{Liouabst}) 
  with $\bra{\Pi(N)}$ from the left.
In case of the second equation we multiply it by $\ket{\Pi(N)}$ from the right however now we have to be more careful.
 On the left side we get
\begin{equation}\label{parpsikom}
	i\frac{d\bra{\psi_N}}{dt}\ket{\Pi_N}=
	i\partial_t \psi_N^*(\Pi)
\end{equation}
using 
\begin{equation}
	\psi_N^*(\Pi)=\left<\Pi(N)|\psi_N\right>^*=\left<\psi_N|\Pi(N)\right> \ . 
\end{equation}
On the other hand if we multiply the right side of the second equation in (\ref{Liouabst}) we obtain 
\begin{eqnarray}\label{psiNcom}
&&-	\bra{\psi_N}\hL_N\ket{\Pi(N)}=
-\left<\hL^\dag_N\psi_N|\Pi(N)\right>=\nonumber \\
-
&&\left<\Pi(N)|\hL_N\psi_N\right>^*=-\hL_N^*(\Pi)\psi^*_N(\Pi)=
\hL_N(\Pi)\psi_N^*(\Pi)	 \ , 
\end{eqnarray}
where we firstly use obvious generalization of (\ref{Ldag}) to $N-$particle case and also the fact that  $\hL_N^*(\Pi)=-\hL_N(\Pi)$.
Then if we combine (\ref{parpsikom}) together with (\ref{psiNcom}) we obtain 
the desired result
\begin{equation}
	i\partial_t \psi_N^*=\hL_N(\Pi)\psi_N^* \ . 
\end{equation}
Now with the help of abstract bra-$\bra{\psi_N}$ and ket-
$\ket{\psi_N}$ vectors we  propose density matrix operator for $N-$particle 
system in the form 
\begin{equation}
	\hrho_N=\ket{\psi_N}\bra{\psi_N}
\end{equation}
that using equations above obeys equation of motion 
\begin{equation}\label{eqrhoN}
i\frac{d\hrho}{dt}=\hL_N \hrho_N-\hrho_N \hL_N \ .
\end{equation}
As a check of our proposal  let us firstly write $\hrho_N$ in a coordinate representation. Using partition of unity in the form 
\begin{equation}
	\hat{\mathrm{I}}_N=\int \prod_{i=1}^N d\Pi_i \ket{\Pi(N)}\bra{\Pi(N)}
\end{equation}
we get
\begin{eqnarray}
	\hat{\rho}_N=\int \prod_{i=1}^N d\Pi_i
\prod_{j=1}^N d\Pi'_j \ket{\Pi(N)}\left<\Pi(N)|\psi_N\right>
\left<\psi_N|\Pi'(N)\right>\bra{\Pi'(N)}
\nonumber \\
\end{eqnarray}
so that
\begin{eqnarray}\label{hatrhocor}
&&	\hat{\rho}_N
=\int \prod_{i=1}^Nd\Pi_i \prod_{j=1}^N
d\Pi'_j \ket{\Pi(N)}\bra{\Pi'(N)}\rho_N(\Pi,\Pi') \ , \nonumber \\ 
&&\rho_N(\Pi,\Pi')=\psi_N(\Pi)\psi^*_N(\Pi')=
\left<\Pi(N)|\hrho_N |\Pi'(N)\right> \  \nonumber \\
\end{eqnarray}
and hence we see that diagonal components of density matrix $\rho_N(\Pi,\Pi)$ exactly correspond to $N-$particle distribution function how is defined in KvN theory
\begin{equation}
	\rho_N(\Pi,\Pi)=\psi_N(\Pi)\psi_N(\Pi)=\rho_N(\Pi) \ . 
\end{equation}
As a next check of our proposal let us multiply (\ref{eqrhoN}) with $\bra{\Pi(N)}$ from the left and $\ket{\Pi'(N)}$ from the right to get
\begin{eqnarray}\label{eqrhoNcor}
	i \partial \rho_N(\Pi,\Pi')=
\hL_N(\Pi)\rho_N(\Pi,\Pi')+\hL_N(\Pi')\rho_N(\Pi,\Pi') 
\end{eqnarray}
using
\begin{eqnarray}
&&\left<\Pi(N)|\hrho_N\hL_N|\Pi'(N)\right>=\left<\Pi'(N)|\hL_N\hrho_N|\Pi(N)\right>^*=\nonumber \\
&&=\hL^*_N(\Pi')\rho_N^*(\Pi',\Pi)=
-\hL_N(\Pi')\rho_N(\Pi,\Pi') \ , \nonumber \\
\end{eqnarray}
and also the fact that
\begin{equation}
	\rho_N(\Pi',\Pi)=\psi_N(\Pi')\psi^*_N(\Pi) \ , 
	\quad 
\rho_N(\Pi',\Pi)^*=\psi_N^*(\Pi')\psi_N(\Pi)=\rho_N(\Pi,\Pi') \ . 
\end{equation}
Note that in the first step we used the fact that
 $\hL_N^\dag=\hL$ and  $\hrho_N^\dag=\hrho_N$ together with $(\hrho_N\hL_N)^\dag=\hL_N^\dag \hrho_N^\dag$. 
It is also easy to see  that (\ref{eqrhoNcor}) in the limit $\Pi\rightarrow \Pi'$ implies following equation
\begin{equation}
	\partial_t
\rho_N(\Pi,\Pi)=\hL_N(\Pi)\rho_N(\Pi,\Pi)
\end{equation}
which is Liouville equation for $N-$particle distribution function.

Main motivation for introduction of the concept of density matrix operator
in KvN theory is that it naturally allows us to 
 define reduced density matrix $\hat{\rho}_n$ by trace over
$N-n$ degrees of freedom 
\begin{equation}
	\hat{\rho}_n\equiv \tr_{N-n}\hat{\rho}_N \ . 
\end{equation}
To do this we write the coordinate basis of the Hilbert space $\ket{\Pi(N)}$ as tensor product of basic vectors of two subspaces 
\begin{eqnarray}
&&	\ket{\Pi(N)}=\ket{\pi_n}\otimes \ket{\sigma}_{N-n} \ , \nonumber \\
&&	\ket{\pi_n}\equiv \ket{\Pi_1}\otimes \ket{\Pi_2}\otimes \dots
	\otimes \ket{\Pi_n} \ , \quad \ket{\sigma_{N-n}}\equiv
\ket{\Pi_{n+1}}\otimes \ket{\Pi_{n+2}}\otimes \dots\otimes \ket{\Pi_N} \ 
\nonumber \\	
\end{eqnarray}
and also use notation
\begin{equation}
	\prod_{i=1}^Nd\Pi_i\equiv d\pi_n d\sigma_{N-n} \ , \quad 
	d\pi_n\equiv \prod_{i=1}^n d\Pi_i \ , 	\quad 
	d\sigma_{N-n}\equiv\prod_{i=n+1}^N d\Pi_i \ . 
\end{equation}
Using these formulas in (\ref{hatrhocor})  we obtain 
\begin{eqnarray}
&&	\hat{\rho}_n=\int d\sigma_{N-n}
 \bra{\sigma_{N-n}}\hat{\rho}_N\ket{\sigma_{N-n}}=\nonumber \\
&& =\int d\sigma_{N-n}d\pi'_n d\sigma'_{N-n}  d\pi''_n d\sigma''_{N-n} \ket{\pi'_n}
 \left<\sigma_{N-n}|\sigma_{N-n}'\right>\times \nonumber \\
 &&\times \psi^*_N(\pi'_n,\sigma'_{N-n})\psi_N(\pi''_n,\sigma''_{N-n})\left<\sigma''_{N-n}|\sigma_{N-n}\right>\bra{\pi''_n}=
 \nonumber \\
 &&=\int d\pi'_n d\pi''_n d\sigma_{N-n}
 \psi_N^*(\pi'_n,\sigma_{N-n})\psi_N(\pi''_n,\sigma_{N-n})\ket{\pi'_n}\bra{\pi''_n} \ , 
 \nonumber \\
\end{eqnarray}
using the fact that 
\begin{equation}
	\left<\sigma_{N-n}|\sigma'_{N-n}\right>=
\prod_{i=n+1}^N\delta(\Pi_{i}-\Pi'_{i})	\ . 
\end{equation}
We see that  $n-$particle density matrix operator has the form  
\begin{equation}
	\hat{\rho}_n=\int d\pi'_n d\pi''_n
	\ket{\pi'_n} \rho_n(\pi',\pi'')\bra{\pi''_n} \ , 
\end{equation}
where coordinate representation of $n-$particle density matrix $\rho_n(\pi'_n,\pi''_n)$ is equal to 
\begin{equation}
	\rho_n(\pi'_n,\pi''_n)=\int \prod_{i=n+1}^N
	d\Pi_i\psi_N^*(\pi'_n,\Pi_{n+1},\dots,\Pi_N)
\psi_N(\pi''_n,\Pi_{n+1},\dots,\Pi_N) \ .
\end{equation}
As the next step we would like to determine equations of motion for reduced density 
matrix $\rho_n(\pi'_n,\pi''_n)$. First of all we have 
\begin{eqnarray}\label{partrho}
&&	\partial_t \rho_n(\pi'_n,\pi''_n)=
	\int d\sigma_{N-n}[\partial_t \psi_N^*(\pi'_n,\sigma_{N-n})
	\psi_N(\pi''_n,\sigma_{N-n})+\nonumber \\
&&+\psi_N^*(\pi''_n,\sigma_{N-n})\partial_t\psi_N(\pi''_n,\sigma_{N-n})] \ ,
\nonumber \\ 
\end{eqnarray}
where 
\begin{eqnarray}
&&	\partial_t\psi_N(\pi'_n,\sigma_{N-n})=\hL_N(\pi'_n,\sigma_{N-n})\psi_N(\pi'_n,\sigma_{N-n})=
\nonumber \\
&&=	\hL_n(\pi'_n)\psi_N(\pi'_n,\sigma_{N-n})+\hL_{N-n}(\sigma_{N-n})\psi_N(\pi'_n,\sigma_{N-n})+
	\hL_{int}\psi_N(\psi'_n,\sigma_{N-n})	\ , 
		\nonumber \\
\end{eqnarray}
where we implicitly presumed that Liouville operator splits into three parts: $\hL_n(\pi_n)$ depends on $\pi$ variables only while $\hL_{N-n}(\sigma_{N-n})$ is function of $\sigma_{N-n}$ variables and finally  $\hL_{int}$  is defined on the whole phase space. In order to find explicit form of these operators we return to  the Hamiltonian for $N-$particle system (\ref{HN}) 
that  can be written as  
\begin{eqnarray}
&&H_N=H_n+H_{N-n}+H_{int} \ ,\nonumber \\
&&H_n=\sum_{\alpha=1}^n\frac{\bp_\alpha\cdot\bp_\alpha}{2m}+V_n \ , \quad  V_n=\frac{1}{2}\sum_{\alpha,\beta=1}^n
V_{\alpha\beta}(\bq_\alpha,\bq_\beta) \ , \nonumber \\
&&H_{N-n}=
\sum_{i=n+1}^N\frac{\bp_i\cdot\bp_i}{2m}+V_{N-n} \ , \quad V_{N-n}=\frac{1}{2}\sum_{i,j=n+1}^N
V_{ij}(\bq_i,\bq_j) \ , \nonumber \\
&&H_{in}=\frac{1}{2}\sum_{i=n+1}^N\sum_{\alpha=1}^n V_{i\alpha}(\bq_i,\bq_\alpha)+
\frac{1}{2}\sum_{\beta=1}^n\sum_{j=n+1}^NV_{\beta j}(\bq_\beta,\bq_j) \ . \nonumber \\
\end{eqnarray}
Then it is easy to see that $\hL_N$ splits in the following way
\begin{eqnarray}
&&	\hL_N=\hL_n+\hL_{N-n}+\hL_{int} \ , \nonumber \\
&&	\hL_n(\pi_n)=i\sum_{\alpha=1}^n
\left(\frac{\partial H_n}{\partial \bq_\alpha}\cdot\frac{\partial}{\partial \bp_\alpha}-
	\frac{\partial H_n}{\partial \bp_\alpha}\cdot\frac{\partial}{\partial \bq_\alpha}\right) \ , \nonumber \\
&&\hL_{N-n}(\sigma_{N-n})=i\sum_{i=n+1}^N
\left(\frac{\partial H_{N-n}}{\partial \bq_i}\cdot\frac{\partial}{\partial \bp_i}-
\frac{\partial H_{N-n}}{\partial \bp_i}\cdot\frac{\partial}{\partial \bq_i}\right) \ , \nonumber \\
\end{eqnarray}
and finally 
\begin{eqnarray}
&&\hL_{int}
=\frac{1}{2}\sum_{\gamma=1}^n\sum_{i=n+1}^N(\frac{\partial V_{i\gamma}}{\partial \bq_\gamma}+
\frac{\partial V_{\gamma i}}{\partial \bq_\gamma})\cdot\frac{\partial}{\partial \bp_\gamma}+
\nonumber \\
&&+\frac{1}{2}\sum_{m=n+1}^N
\sum_{\alpha=1}^n(\frac{\partial V_{i\alpha}}{\partial \bq_m}+\frac{\partial V_{\alpha m}}{\partial \bq_m}) \cdot\frac{\partial}{\partial \bp_m} \ . \nonumber \\
	\end{eqnarray}
Let us now discussion contribution of each $\hL'$ in (\ref{partrho}) separately.	
In case of $\hL_n(\pi_n)$ we obtain following
result
\begin{eqnarray}
&&		\int d\sigma_{N-n}[
(\hL_n(\pi'_n)\psi_N^*(\pi'_n,\sigma_{N-n})\psi_n(\pi''_n,\sigma_{N-n})+\psi_N^*(\pi'_n,\sigma_{N-n})\hL_n(\pi''_n)\psi_N(\pi''_n,\sigma_{N-n})=\nonumber \\
&&=\hL_n(\pi'_n)\rho_n(\pi'_n,\pi''_n)+\hL_n(\pi''_n)\rho_n(\pi'_n,\pi''_n)		
	\end{eqnarray}
since $\hL_n(\pi'_n),\hL_n(\pi''_n)$ do not depend on integration variables. 
On the other hand in case of $\hL_{N-n}(
\sigma_{N-n})$ 	we obtain
\begin{eqnarray}\label{sigmaNn}
&&\int d\sigma_{N-n}(\hL_{N-n}(\sigma_{N-n})\psi^*_N(\pi'_n,\sigma_{N-n})\psi_N(\pi''_n,\sigma_{N-n})+\nonumber \\
&&+
\psi^*_N(\pi'_n,\sigma_{N-n})\hL_{N-n}(\sigma_{N-n})\psi_N(\pi''_n,\sigma_{N-n}))=\nonumber \\
&&=\int d\sigma_{N-n}\hL_{N-n}(\psi_N^*(\pi'_n,\sigma_{N-n})\psi_N(\pi''_n,\sigma_{N-n}))=0
\end{eqnarray}
since this expression contains derivative with respect to $p_i$ or $q_i,i=n+1,\dots,N$ that thanks to the integration over these variables gives contributions  of $\psi_N$ at asymptotic region where we can impose  appropriate boundary conditions so that 
(\ref{sigmaNn}) vanishes.

Finally we discuss  contribution  of $\hL_{int}$ in time evolution 
$\rho_n(\pi'_n,\pi''_n)$.
Let us now presume that the potential $V_{ij}$ depends on distance 
$d_{ij}\equiv |\bq_i-\bq_j|^2$. Then clearly $d_{ij}=d_{ji}$ and $V_{ij}=V_{ji}$ so that 
the interaction operator simplifies as 
\begin{equation}\label{hLint}
	\hL_{int}=\sum_{\gamma=1}^n\sum_{i=n+1}^N
	\frac{\partial V_{i\gamma}}{\partial \bq_\gamma}\cdot\frac{\partial }{\partial \bp_\gamma}+
	\sum_{m=n+1}^N\sum_{\alpha=1}^n
	\frac{\partial V_{i\alpha}}{\partial \bq_m}\cdot\frac{\partial}{\partial \bp_m} \ . 
\end{equation}
Now it is clear that the second term in (\ref{hLint}) gives zero contribution in the equation of motion for $\rho_n(\pi'_n,\pi''_n)$ using the same arguments as in case of operator $\hL_{N-n}(\sigma_{N-n})$.
 Then let us concentrate on the first term in (\ref{hLint})  and restrict
to the special case $n=N-2$. Then we have
\begin{eqnarray}
&&\int d\sigma_1 d\sigma_2\sum_{\gamma=1}^{N-2}\sum_{i=1}^2
\frac{\partial V_{i 1}}{\partial \bq_\gamma}\cdot\frac{\partial}{\partial \bp'_\gamma}\psi^*_N(\pi'_n,\sigma_{1},\sigma_{2})\psi_N(\pi''_n,
\sigma_{1},\sigma_{2})=\nonumber \\
&&=\int d\sigma_1 d\sigma_2 \sum_{\gamma=1}^{N-2}(\frac{\partial V_{1\gamma}(\sigma_1,q_\gamma)}{\partial \bq_\gamma}\cdot\frac{
\partial \psi^*_N(\pi'_n,\sigma_1,\sigma_2)}{	
	\partial \bp'_\gamma}\psi_N(\pi''_n,\sigma_1,\sigma_2)+\nonumber \\
&&\frac{\partial V_{2\gamma}(\sigma_2,q_\gamma)}{\partial \bq_\gamma}\cdot\frac{\partial
\psi^*_N(\pi'_n,\sigma_1,\sigma_2)}{\partial \bp'_\gamma}\psi_N(\pi''_n,\sigma_1,\sigma_2)=\nonumber \\
&&=2 \int d\sigma_1 d\sigma_2 (\frac{\partial V_{1\gamma}(\sigma_1,\bq_\gamma)}{\partial \bq_\gamma}\cdot \frac
{\partial \psi^*_N(\pi'_n,\sigma_1,\sigma_2)}
{\partial \bp'_\gamma}\psi_N(\pi''_n,\sigma_1,\sigma_2)=\nonumber \\
&&=2\int d\sigma \sum_{\gamma=1}^{N-2}
\frac{\partial V_{\sigma\gamma}(\sigma,g_\gamma)}{
\partial \bq_\gamma}\cdot\frac{\partial}{\partial \bp'_\gamma}\rho_{N-1}(\pi'_n,\sigma,\pi''_n,\sigma) \ , 
\nonumber \\
\end{eqnarray}
where we used crucial presumption of  symmetry of the function 
\begin{equation}
	\psi_N(\pi_n,\sigma_1,\sigma_2)=\psi_N(\pi_n,\sigma_2,\sigma_1)
\end{equation}
and where we defined $\rho_{N-1}(\pi',\sigma',\pi'',\sigma'')$ as
\begin{equation}
\rho_{N-1}(\pi'_n,\sigma',\pi''_n,\sigma'')=
\int d\sigma_1 \psi_N^*(\pi'_n,\sigma',\sigma_1)\psi_N(\pi''_n,\sigma'',\sigma_1) \ . 
\end{equation}
In the same way we can manipulate with $\hL_{int}$ acting on $\psi_N$ and we obtain final result
\begin{eqnarray}
&&	\partial_t \rho_{N-2}(\pi'_n,\pi''_n)=
	\hL_{N-2}(\pi'_n)\rho_{N-2}(\pi'_n,\pi''_n)+
	\hL_{N-2}(
	\pi''_n)\rho_{N-2}(\pi'_n,\pi''_n)+\nonumber \\
&&+2\sum_{\gamma=1}^{N-2}\int d\sigma\left[ 
\frac{\partial V_{\sigma \gamma}(\sigma,q'_\gamma)}{
\partial \bq'_\gamma}\cdot\frac{\partial}{\partial  \bp'_\gamma}\rho_{N-1}(\pi'_n,\sigma,\pi''_n,\sigma)+
\frac{\partial V_{\sigma \gamma}(\sigma,q''_\gamma)}{
	\partial \bq''_\gamma}\cdot\frac{\partial}{\partial \bp''_\gamma}\rho_{N-1}(\pi'_n,\sigma,\pi''_n,\sigma)\right] \ . 
\nonumber \\
\end{eqnarray}
Since $2=N-(N-2)=N-n$ we see that this result can be easily
generalized to the case of $n-$particle density matrix and we obtain following equation 
\begin{eqnarray}
&&	\partial_t \rho_{n}(\pi'_n,\pi''_n)=
	\hL_{n}(\pi'_n)\rho_{n}(\pi'_n,\pi''_n)+
	\hL_{n}(
	\pi''_n)\rho_n(\pi'_n,\pi''_n)+\nonumber \\
&&	+(N-n)\sum_{\gamma=1}^{n}\int d\sigma\left[ 
	\frac{\partial V_{\sigma \gamma}(\sigma,\bq'_\gamma)}{
		\partial \bq'_\gamma}\cdot \frac{\partial}{\partial  \bp'_\gamma}\rho_{n+1}(\pi'_n,\sigma,\pi''_n,\sigma)+
	\frac{\partial V_{\sigma \gamma}(\sigma,q''_\gamma)}{
		\partial \bq''_\gamma}\cdot\frac{\partial}{\partial \bp''_\gamma}\rho_{n+1}(\pi'_n,\sigma,\pi''_n,\sigma)\right] \ , 
	\nonumber \\
\end{eqnarray}
where we again defined $\rho_{n+1}(\pi'_n,\sigma',\pi''_n,\sigma'')$ as
\begin{equation}
\rho_{n+1}(\pi'_n,\sigma',\pi''_n,\sigma'')=
\int \prod_{i=n+2}^N d\Pi_i
\psi^*_N(\pi'_n,\sigma',\Pi_{n+2},\dots,\Pi_N)
\psi_N(\pi''_n,\sigma'',\Pi_{n+2},\dots,\Pi_N) \ .
\end{equation}
The last expression is final result of our paper and it is simply 
analogue of BBGKY hierarchy
\cite{Yvon,Bogoliubov,Kirkwood}
 for density matrix of KvN formulation of N-particles. 
Of course, standard BBGKY equation arises in the limit when $\pi'_n\rightarrow \pi''_n$.
\section{Conclusion}\label{fourth}
Let us outline our results and suggest possible extension of this work. We studied KvN theory for $N-$particle system and we argued that relation between $N-$particle distribution function and corresponding wave function $\psi_N$ as was proposed in KvN theory should be interpreted as diagonal components of density matrix operator in coordinate representation. Then we proposed corresponding density matrix operator in terms of abstract bra-ket notation and determined its equation of motion.  Then we naturally defined  reduced density matrix operator  
when we performed the trace of $N-$particle density operator over subspace orthogonal to $6n-$dimensional phase space. Finally we 
determined equations of motion for reduced density matrix $\rho_n$ which is analogue of BBGKY hierarchy for distribution functions in statistical mechanics. 

In our opinion an introduction of density matrix operator in operational formulation of classical physics is non-trivial step and offers new opportunities to study relation between classical and quantum physics. For example, it would be interesting to study an analogue of entanglement in classical physics using  this density matrix operator. Further, since density  matrix in KvN theory  is given as a product of two wave functions it corresponds to the pure state. It would be interesting whether it is possible to generalize it to mixed states and 
study its  physical meaning. We hope to return to these problems in future.   
\\
\\
{\bf Acknowledgment:}
This work  was
supported by the Grant Agency of the Czech Republic under the grant
GA26-22343S.

\end{document}